%% file: main.tex
\documentclass[11pt]{article}
\usepackage[left=2.5cm, right=2.5cm, top=2.5cm, bottom=2.5cm]{geometry}

\usepackage{mathptmx}
\DeclareMathAlphabet{\mathcal}{OMS}{cmsy}{m}{n}

\usepackage{xcolor}
\usepackage{amsmath}
\usepackage{amssymb}
\usepackage{mathtools}
\usepackage{hyperref}
\usepackage{graphicx}
\usepackage{subfig}
\usepackage{caption}
\usepackage{titlesec} 
\usepackage{enumitem} 
\usepackage{fancyhdr}

\usepackage[noend]{algpseudocode}
\usepackage{algorithm}
\usepackage{algorithmicx}
\makeatletter
\renewcommand{\@seccntformat}[1]{}
\makeatother

\titlespacing{\section}{0pt}{-0.1em}{-\parskip}

\titlespacing{\subsection}{0pt}{0.1em}{-\parskip}

\setitemize{noitemsep, topsep=-1em, parsep=0pt, partopsep=0pt}

\newcommand{\myfontsize}[2]{{\fontsize{#1}{0}\selectfont#2}}

\DeclareMathOperator*{\argmax}{argmax} 
\renewcommand{\S}{\mathcal{S}}
\newcommand{\A}{\mathcal{A}}
\newcommand{\R}{\mathcal{R}}
\newcommand{\E}{\mathbb{E}}
\newcommand{\MUBEV}{MUBEV-SM}
\newcommand{\doubleplus}{+\!\!\!+}

\algrenewcommand\algorithmicrequire{\textbf{Input:}}
\algrenewcommand\algorithmicensure{\textbf{Output:}}

\DeclareCaptionLabelFormat{mylabel}{#1 #2 - }

\renewenvironment{thebibliography}[1]{
    \begin{oldthebibliography}{#1}
        \setlength{\itemsep}{0em}
        \setlength{\parskip}{0em}
    }
    {
    \end{oldthebibliography}
}

\usepackage{lipsum}

\fancypagestyle{firstpage}
{
    \pagestyle{fancy}
    \fancyhead[L]{\centering{27$^\text{th}$ ITS World Congress, Hamburg, Germany, 11-15 October 2021}}    
    \fancyhead[R]{}
}
\begin{document}
    \captionsetup[figure]{
        labelfont={small,bf},
        labelformat=mylabel,
        labelsep=none,
        name=Figure,
        font={small,bf}
    }
    \thispagestyle{firstpage}
    \begin{center}
        \myfontsize{14}{Paper ID \#889}\\

        \myfontsize{14}{\textbf{
            Spatial Positioning Token (SPToken) for Smart Parking
        }}
        \\
        \medskip

        \textbf{
            Roman Overko$^{\text{1*}}$,
            Rodrigo Ord\'{o}\~{n}ez-Hurtado$^\text{2}$,
            Sergiy Zhuk$^\text{2}$,
            Robert Shorten$^\text{3}$
        }
        \\
        \myfontsize{10}{
            1. University College Dublin, Dublin, Ireland.
            Email \href{mailto:roman.overko@ucdconnect.ie}{roman.overko@ucdconnect.ie}\\
            2. IBM Research, Damastown Industrial Park, Dublin 15, Ireland\\
            3. Imperial College London, South Kensington, London SW7 2AZ, United Kingdom\\
        }
    \end{center}
    \textbf{Abstract}\\
        In this paper, we describe an approach to guide drivers searching for a parking 
        space (PS). The proposed system suggests a sequence of routes that drivers should 
        traverse in order to maximise the expected likelihood of finding a PS and minimise the
        travel distance. This system is built on our recent architecture SPToken, which combines
        both Distributed Ledger Technology (DLT) and Reinforcement Learning (RL) to realise a 
        system for the estimation of an unknown distribution without disturbing the environment. 
        For this, we
        use a number of virtual tokens that are passed from vehicle to vehicle to enable a massively
        parallelised RL system that estimates the best route for a given origin-destination (OD) pair, 
        using crowdsourced information from participant vehicles. Additionally, a moving 
        window with reward memory mechanism is included to better cope with non-stationary 
        environments. Simulation results are given to illustrate the efficacy of our system.


    \section{\myfontsize{11}{Introduction}}
    \input{01_introduction.tex}

    \section{\myfontsize{11}{Related work}}
    \input{02_related_work.tex}

    \section{\myfontsize{11}{SPToken for Smart Parking}}
    \label{sec:sptoken}
    \input{03_00_sptoken.tex}

    \subsection{\myfontsize{11}{\textnormal{\textit{Optimal policy search algorithm}}}}
    \label{sec:algorithm}
    \input{03_03_algorithm.tex}

    \subsection{\myfontsize{11}{\textnormal{\textit{Reward function}}}}
    \input{03_04_reward_func.tex}


    \section{\myfontsize{11}{Numerical evaluation}}
    \input{04_evaluations.tex}

    \section{\myfontsize{11}{Conclusion}}
    \input{05_conclusions.tex}

    \section{\myfontsize{11}{Acknowledgements}}
        This work has been supported by Science Foundation Ireland (SFI) under grant No. 16/IA/4610.
        Part of this work has received funding from the EU's Horizon 2020 research \& innovation 
        programme via the ICT4CART project under grant agreement No. 768953. 
        Content reflects only the authors' view and the European Commission is 
        not responsible for any use that may be made of the information it contains.

    \bibliographystyle{ieeetr}
    \bibliography{refs}
\end{document}

%% file: 01_introduction.tex

A problem of immense importance to drivers is that of finding an available parking space (PS), which is also a societal problem affecting
air pollution, congestion, and noise emission due to the large number of vehicles currently on the streets. For illustration, a case study
of a small business district in Los Angeles \cite{shoup2006cruising} indicated that 730 tons of CO$_2$ where produced and
47,000 gal of gasoline were burned in a year by cars searching for parking. Another study by
McKinsey\footnote{\scriptsize{\url{https://www.mckinsey.com/industries/public-and-social-sector/our-insights/the-smart-city-solution}}} 
reported that the average car owners in Paris spend about four years of their lives searching for PSs.
The emerging paradigm of intelligent transport systems (ITS) offers a great hope to alleviate these problems by enabling a range of
parking guidance systems that are rapidly becoming an essential part of future sustainable cities. The most common approach
are information boards displaying the available PSs at various locations around the city. This system provides a valuable guidance for
drivers to avoid areas with potentially limited PS availability (PSA), but often leads to localised congestion around areas 
with the largest number of available PSs which is caused by (i) all the drivers receiving identical data, and (ii) the trend of most drivers to
typically choose the areas with large PSA. More modern systems~\cite{fog} provide extensive parking information directly to
the drivers either to their smart devices or to the cars.
The wealth of PSA information is provided to the drivers through the use of
PS sensors. Such systems thus provide drivers with the exact location of currently free PSs along with e.g. their respective prices.
The main disadvantage of such systems is that a large investment cost is required as sensors are expensive to deploy and maintain,
thus limiting large-scale deployments. On the other hand, if moving vehicles are used to collect raw environment data (e.g. from dash
cams or LiDAR sensors) to be turned into PSA information in cloud/MEC servers, then the cost and complexity of maintaining a central database
of parking information built from raw environment data is also prohibitive from the perspective of the burden on the telecommunications network.

In the context of the EU, the European Commission has created a platform for the deployment of Cooperative ITS, namely the C-ITS
Platform\footnote{\scriptsize{\url{https://ec.europa.eu/transport/themes/its/c-its_en}}}, with the aim of improving the efficiency of road
transport at European-wide scale through a number of identified (essential) applications including {\it probe vehicle data} and {\it parking
services}. In alignment with this, the smart parking solution proposed in this paper aims to use moving vehicles as collectors and processors of raw
environment data so that PSA can be shared as floating car data. For instance, if the ICT architecture proposed in~\cite{ict4cart2020}
is integrated with the Spatial Potisioning Token (SPToken) framework \cite{sptoken_journal, iccve}, then SPToken observers could include
connected automated vehicles (CAVs), road-side units (RSUs), or MEC servers, and the floating car data can be transmitted via (perhaps
extended) Cooperative Awareness Messages and Collective Perception Messages \cite{garlichs2020leveraging} (or a new message type to be developed).


This paper is organised as follows. First, we provide a brief summary of related work on applications in the area of smart parking.
Second, we proceed to describe in detail the integration of SPToken into a smart parking solution. Third, we provide numerical validation of the
proposed approach. Finally, we close the paper with some concluding remarks.


%% file: 02_related_work.tex

We proceed to present a short overview of relevant smart parking solutions. In~\cite{fog}, the authors
propose a MEC-based architecture using VANETs to improve the parking experience concerning
average parking cost, fuel wastes, and vehicle exhaust emissions, and the solution gives drivers the 
option to integrate their own parking preferences. However, this solution relies on parking information from
interconnected MEC servers deployed at parking lots, thus substantially restricting its scope to such
locations and not allowing for the integration of unsupervised on-street parking.
Concerning privacy-preserving data management, several architectures for smart parking solutions have
already been proposed. In~\cite{blockchain_parking}, a blockchain-based descentralised parking management service is designed, which
ensures anonymous authentication and resistance to data linkability, among others.
In~\cite{p_span}, the authors introduce a privacy-preserving smart parking navigation system based on
the integration of cloud services and vehicular communication, that allows for identity/service authentication
and traceability. Even though \cite{blockchain_parking} and \cite{p_span} provide means for parking data retrieval
(including PSA and navigation guidance) or PS reservation, their allocation process relies on parking-related
databases maintained with availability data provided by e.g. parking lot owners, and it does not involve any optimisation problem.
The literature on applications of machine learning to smart parking is vast and we simply refer to a few 
relevant publications. In~\cite{rl_lavp}, the authors propose a Q-learning-based solution to 
find the nearest PS by minimising the total covered distance, time taken and 
consumed energy. This approach is evaluated only on stationary environments, which is restrictive in many 
real-world applications including traffic management problems. Situations with non-stationary
models may lead to suboptimal policies provided by RL algorithms. In this context, only a few existing 
works have considered to improve RL algorithms for non-stationary environments. A recent example of such 
works includes~\cite{non_stat_rl}, in which the authors have developed a model-free RL method to effectively detect 
changes in the rewards and transition dynamics, and validated the proposed approach on randomly generated data
and traffic signal control problems as well.


%% file: 03_00_sptoken.tex

In this work, we make use of the  Distributed Ledger Technology (DLT) based SPToken architecture proposed in~\cite{sptoken_journal, iccve} to
design a solution for the on-street parking problem. As detailed in~\cite{sptoken_journal, iccve}, the SPToken framework allows to explore a given
environment without perturbing it, this achieved through the use of (i) virtual entities referred to as {\it tokens} in combination with (ii) crowdsourcing
approaches such as multi-agent RL, which makes SPToken suitable for a wide range of smart mobility applications including smart parking.
The key idea is to use the tokens as virtual containers to be filled with floating car data, and allow tokens to ``jump'' from vehicle to vehicle whenever
is required to complete routes determined by the underlying RL algorithm. A participating vehicle can ``collect'' a free token via a DLT transaction if its route 
coincides with the token route, after which the vehicle with the token turns into an RL agent
that updates the distributed ledger with some information (e.g. spent travel time, 
measured roadside parking availability) whenever it passes an {\it observer}. The vehicle must ``deposit''
the token via another DLT transaction when it deviates from the token route.
Since tokens are transported by participating vehicles (not forced to follow the entire token route) already present in an urban scenario, SPToken guarantees that the environment stays undisturbed
during the probing process. Note that the physical presence of vehicles is imperative to collect/deposit tokens,
and so a {\it Proof-of-Position} mechanism is also included (see \cite{iccve} for details).
SPToken also has all the advantages of DLT such as data privacy preservation, data ownership retention, and misuse/spamming prevention.
As a base requirement for participating vehicles, we assume they can provide on-street PSA data
along their routes as a result of applying analytics to (raw) real-time environment data collected by their on-board sensors
(e.g. point clouds from LiDAR sensors). We assume this requirement is reasonably satisfied based on reported mechanisms for on-board
detection of parking availability \cite{wu2019early, bock2015street}.

Note that the proposed solution is not a straightforward
application of the SPToken architecture to smart parking, and a number of extensions (also to the underlying RL algorithm~\cite{sptoken_journal}) were required:
(i) a new mechanism called \textit{moving window with reward memory} (MWRM) for MUBEV~\cite{iccve} to
improve the performance in non-stationary environments;
(ii) a more advanced action selection strategy to increase exploration at early stages of the learning process;
(iii) an improved design of the reward function for the RL algorithm to account for a reduced number of tuning parameters; 
and (iv) a design parameter $\alpha$ is introduced as a means for users to provide their preferences in terms of the their preferred optimisation objectives.
Additionally, the application of SPToken to the on-street parking problem required dealing with a non-trivial multi-objective optimisation problem: our solution
involves the investigation of optimal routes connecting origin-destination (OD) pairs so that in such routes (a) travel distance is minimised, and (b) PSA is
maximised. While (a) involves a (widely studied) shortest path (SP) problem,
the solution for (b) requires solving a longest path problem of NP-hard nature.
Therefore, we decided to integrate these two problems into
a graph-based, multi-objective optimisation approach by linearly combining two directed weighted graphs: a travel Distance Graph (DG), and a Parking Availability Graph (PAG).
Both graphs come from the road network, where each vertex is a road junction, each edge is
a collection of roads, weights $\Lambda$ in DG represent the physical (time invariant) length of given roads, and weights $K$ in PAG
are in function of the (time-variant) number of unoccupied on-street PSs along given roads at certain time. As a result, the graph weights
of the convex linear combination $G = (\alpha)\Lambda + (1-\alpha)K$ 
are computed as
\begin{equation}
    \label{alg:weight_func}
    \mathcal{W}_G(\alpha, s, \lambda_s, \nu_s(t)) = 
    \left(\alpha\right)*\lambda_s + \left(1-\alpha\right)*\kappa_s\left(\nu_s(t)\right),
\end{equation}
where: $\alpha$ is a user-defined parameter that satisfies $\alpha_{min} \leq \alpha \leq 1$ and represents the priority given to the distance cost; 
$\lambda_s$ is the length of edge $s$ in G;
$\kappa_s(\nu_s(t)) = -\nu_s(t)*L_{max} / C_{max}$, with $\nu_s(t)$ being the current roadside PSA
along the road links included in edge $s$ at time $t$,
$L_{max}$ the maximum edge length in the road network,
and $C_{max}$ the maximum edge parking capacity in the road network.
While SP routing is widely used as the {\it de-facto} default solution (DS) in most navigation systems, SP does not always guarantees maximum PSA. 
However, the total PS capacity along road links is indeed time-invariant, and thus in our approach we compute DSs using
Equation \ref{alg:weight_func} with the value of $\alpha$ provided by the users. In addition, $\alpha_{min}$ is determined empirically such that the graph
$G$ does not contain negative cycles for weights computed as  $\mathcal{W}_G(\alpha, s, \lambda_s, c_s)$, where $c_s$ is
the total (time-invariant) roadside parking capacity along the road segments included in edge $s$.
Consequently, if $\alpha \in [\alpha_{min}, 1]$, then we can use an SP algorithm to compute DSs to be used as
the initial policy of the RL algorithm. Clearly, DSs degenerate into SPs when $\alpha=1$ (as per Equation~\ref{alg:weight_func}).

%% file: 03_03_algorithm.tex

As previously mentioned, we use an RL strategy to solve our 
target problem. A full RL problem is usually modeled as a Markov Decision Process (MDP). Our decision 
problem is a stationary Finite Horizon MDP (FHMDP), which is a discrete-time 
stochastic control process defined by a tuple $\langle \S, \A, P, \R, H \rangle$, where:
$\S$ is the state space, $\A$ is the action space, $P$ is the tensor of transition probabilities,
$\R$ is the reward matrix, and $H$ is the length of the time horizon. Let $S = |\S|$, $A = |\A|$ and 
$A_s = |\A_s|$, that is, $S$ is the number of states, $A$ is the total number of actions, and $A_s$ is the 
number of actions allowable in state $s \in \S$. We assume that all the states are fully observable. 
We also adopt this general approach: each state $s$ has a different set of allowable actions $\A_s$, hence 
$\A = \bigcup_{s \in \S}\A_s$. Finally, $P(s'|s, a)$ is the probability of transition 
to state $s'$ if action $a$ is chosen in state $s$, and $\R(s, a)$ is the reward of playing action $a$ in 
state $s$.

An RL agent, i.e. the decision maker, interacts with the environment in a sequence of episodes, each 
episode with length $H$. Within an episode $k$ at each time step
$t \in [H]$ (where $[\bullet] = \{1,2,...,\bullet\}$ with $\bullet$ an integer number),
the agent plays 
action $a$ based on the observation of state $s$ and receives a reward $r(s, a)$. The next state 
$s'$ is drawn from a distribution $P$ which defines the trajectory of the MDP, that is 
$s' \sim P(\cdot | s, a)$. Over each episode $k$, the agent selects actions according to a policy $\pi_k$,
which maps states and time steps to actions. $\pi_k$ is updated after $H$ 
interactions with the environment, i.e. at the end of each episode. Note that in stationary MDPs, the 
transition probabilities and reward distribution do not vary with time step $t$. The policy $\pi_k$, 
however, is generally time-step-dependent for FHMDPs. The expected return until the end
of an episode is represented by a \textit{value function} (for state $s$, time step $t$, 
and policy $\pi_k$) defined as
\begin{equation}
    V_{t}^{(\pi_k)}(s_t) \coloneqq \E\left[\sum\nolimits_{i=t}^{H}\R\big(s_i, \pi_k(s_i, i)\big)\right],
    \label{eq:value_func}
\end{equation}
where the expectation is taken with respect to states $s'$ encountered in the MDP. 
The quality of a policy at episode $k$ is characterised by the \textit{total expected reward} defined as
$u^{(\pi_k)} = p_{0}^{\top} V^{(\pi_k)}_1$, 
where $p_0$ is the distribution of the initial states.
The value function represented in Equation~\ref{eq:value_func} can be rewritten as follows:
\begin{equation}
    V_{t}^{(\pi_k)}(s_t) = \R\big(s_t, \pi_k(s_t, t)\big) + P\big(\cdot | s_t, \pi_k(s_t, t)\big)^{\top}
    V_{t+1}^{(\pi_k)}.
    \label{eq:bellman}
\end{equation}
Equation~\ref{eq:bellman} is called the \textit{Bellman equation} or the optimality equation, 
which is often solved using the \textit{backward induction} process with boundary condition 
$V_{H+1}^{(\pi_k)} \coloneqq 0$.
The goal of an RL agent is then to find an optimal trajectory which maximises the total expected 
reward. Thus, the optimal policy $\pi^*$ is calculated through the backward induction procedure 
as follows:
\begin{equation}
        \pi^{*}(s, t) =
            \argmax_{a\in\A_s} \left\{\R(s, a) + P\big(\cdot | s, a\big)^{\top} V_{t+1}^{\pi^*}\right\},
            \,\,\,
        \pi^{*}(s, H) = \argmax_{a\in\A_s} \R(s, a).
    \label{eq:optimal_policy}
\end{equation}

We are now in a position to present the \textbf{M}odified \textbf{UBEV} for \textbf{S}tationary 
FHMDPs with \textbf{M}oving window (\MUBEV) algorithm, which is the tool we use in our approach.
Algorithm~\ref{alg:modified_ubev} represents a new variant of the MUBEV algorithm introduced 
in~\cite{cdc} (see~\cite{sptoken_journal} for the most recent version of MUBEV), which also includes
the key modifications of the UBEV-S algorithm presented in~\cite{pmlr-v80-zanette18a} in order to 
achieve a better performance on stationary MDPs. Namely, UBEV-S incorporates a slightly 
modified exploration bonus and produces tighter regret bounds in the settings of time-independent 
rewards and transition dynamics~\cite{pmlr-v80-zanette18a}.

It is worth noting that \MUBEV\ includes all the modifications introduced in 
the MUBEV algorithm~\cite{iccve, cdc}. That is, (i) \MUBEV\ does not require the estimation of 
transition probabilities due to the specific design of the 
state-action space; (ii) in \MUBEV, observation of the environment is achieved using multi-agent 
approach; and (iii) \MUBEV\ incorporates a better 
action selection strategy. However, we introduce a few new modifications to the \MUBEV\ algorithm as explained below.

First, the action selection strategy is slightly improved now. Specifically, MUBEV is forced to use the 
default policy (e.g. SP policy) whenever all the entries 
of $Q_s$ are equal to each other, where $Q_s$ denotes a vector of $Q$-function values for state $s$.
In \MUBEV, we explicitly compute and use the maximum value of $Q_s$, i.e. $Q_{max}$
(Algorithm~\ref{alg:modified_ubev}, line 14).
When $Q_{max}$ appears multiple times in $Q_s$, the selection strategy is as follows:
if one of the repetitions corresponds to the default action, then action is chosen according to the 
default policy;
otherwise, an entry is randomly selected from those repetitions using uniform distribution, and the
associated action is chosen (Algorithm~\ref{alg:modified_ubev}, line 16).
Such a randomness boosts the 
exploration at the beginning of the learning process.

Second, \MUBEV\ involves a new mechanism not used in~\cite{sptoken_journal, iccve, cdc}, referred to as MWRM,
which is introduced as a mitigation method against stagnation that may occur when a change in the 
environment happens after a long period of stationarity, such as a long sequence of episodes with similar 
traffic conditions.
For the calculation of the estimated reward $\hat{r}$ in UBEV and UBEV-S~\cite{pmlr-v80-zanette18a, ubev}, all reward values from the first episode are 
accumulated, while in \MUBEV\ this only happens in the context of a moving window. That is, in \MUBEV, 
a maximum of past $J$ rewards are accumulated (with $J$ the length of the moving window), and the oldest 
reward is dropped every time a new reward is collected. Additionally, the accumulation is done using
weighted rewards, and all the collected statistics are reset once the confidence bound reaches a certain 
minimum value.
As a result, \MUBEV\ can better cope with environments whose reward distribution dynamically changes or
includes long periods of stationarity\footnote{Note that reward is independent of time step $t$ 
in stationary MDPs, although it can vary from one episode to another.}. However, the MWRM 
mechanism can be disabled if required ($MWRM=\text{False}$ in Algorithm~\ref{alg:modified_ubev}).

\begin{figure}[th]
    \centering
    \subfloat{{\includegraphics[width=0.45\columnwidth]{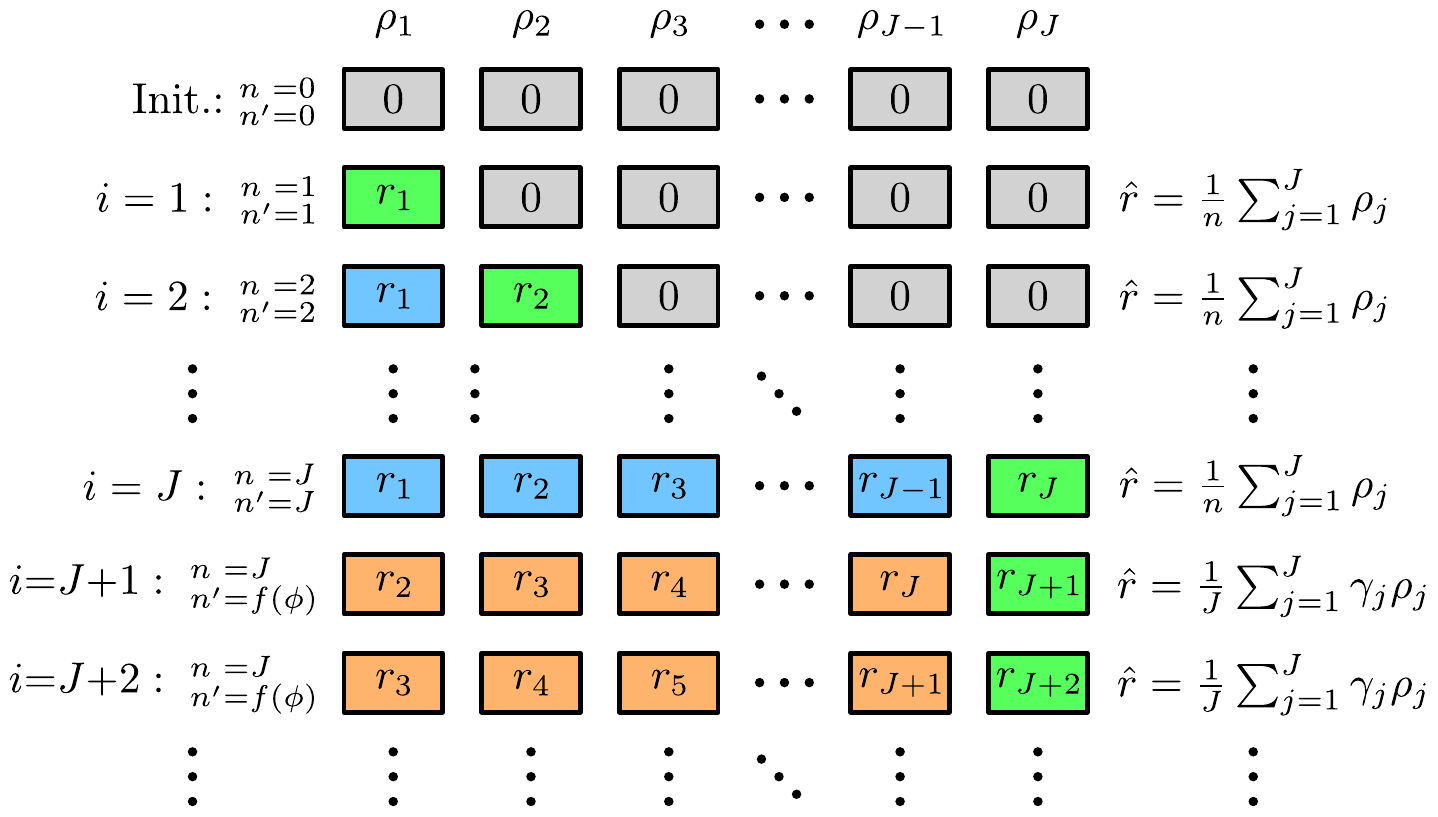}}}
    \subfloat{{\includegraphics[width=0.45\columnwidth]{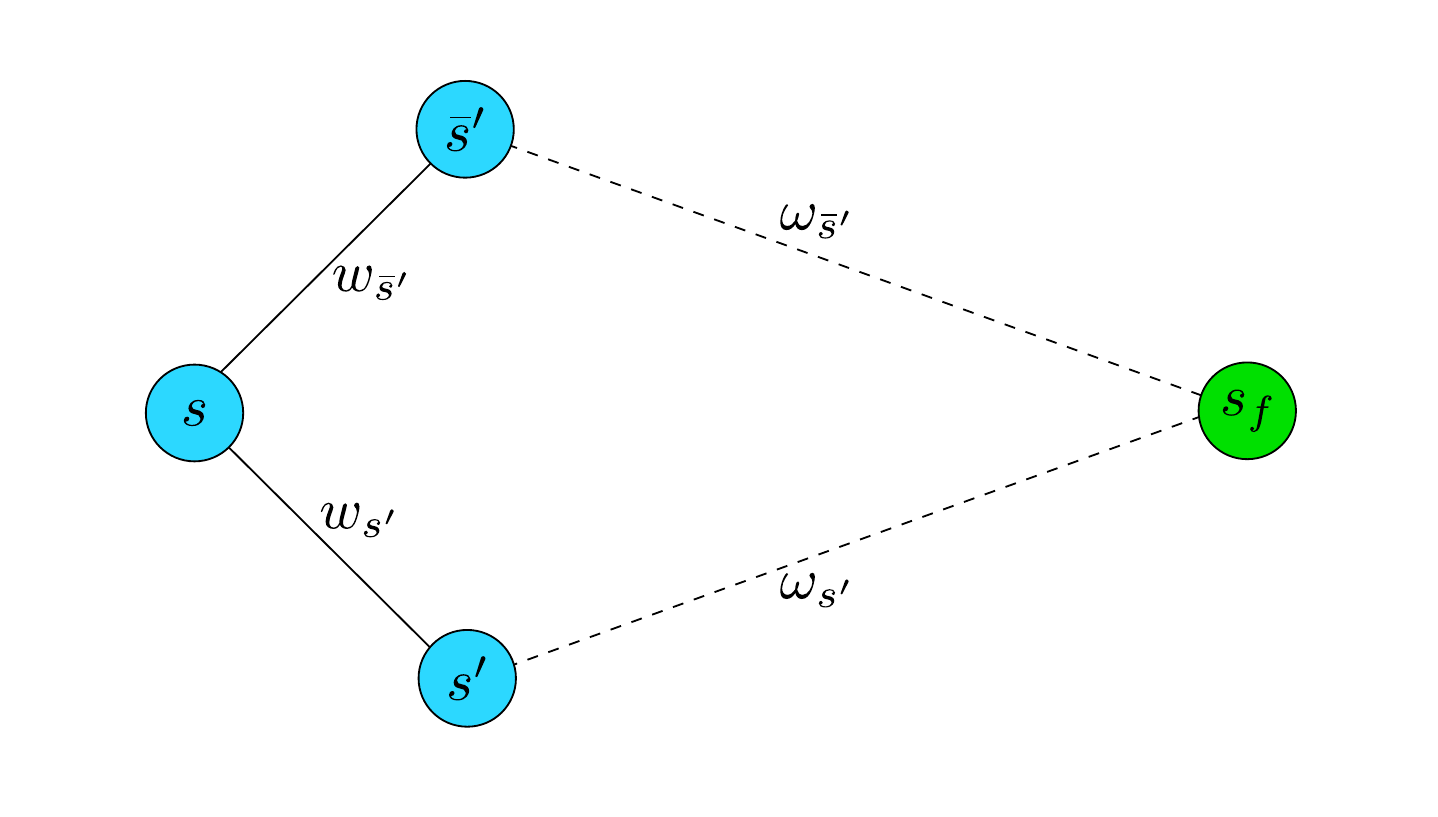}}}
    \caption{
        \underline{Left:} 
        The MWRM mechanism used in \MUBEV. Gray cells represent initial
        values of $\rho$. A green cell represents the current observed reward. Rewards observed at previous 
        interactions are stored sequentially (blue cells). Values of $\rho$ in orange and green cells (if moving 
        window is reached) are subject to discounts.
        \underline{Right:} 
        Schematic representation of the state weight $w$ and the route weight $\omega$
        used for computation of rewards. $w_{s'}$ and $w_{\bar{s}'}$ are the state weights for nodes $s'$ 
        and $\bar{s}'$, respectively. $\omega_{s'}$ and $\omega_{\bar{s}'}$ are the route weights to
        the destination node $s_f$ from nodes $s'$ and $\bar{s}'$, respectively.
    }
    \label{fig:moving_window}
\end{figure}

For a better understanding of the MWRM mechanism, see Figure~\ref{fig:moving_window}-Left in which
we illustrate this technique for a given state-action pair 
$(s, a)$ observed at the $i$-th interaction with the environment (without loss of generality). 
Recall that $J$ is the length of the moving 
window. For each $(s, a)$ pair, we introduce an additional transition count $n'$, and a \textit{reward memory 
vector} $\rho$ of length $J$ to store the collected rewards. Initial values of $n$, $n'$, and $\rho_j$ 
(gray cells) are set to zero 
for all $(s, a)$ pairs and $\forall j \in [J]$, where $n$ and $n'$ represent the number of times action $a$ 
is played in state $s$. Note that both $n$ 
and $n'$ are equal to each other and incremented by $1$ only if $i < J$. At interaction $i$, reward 
$r_i$ is observed and $\rho$ is updated such that $\rho_i = r_i$ (green cells) if $i \leq J$, 
i.e. within the moving window, and the estimated reward $\hat{r}$ is the mean of the collected 
rewards. When the moving window 
length is reached, i.e. $n = n' = J$, we stop incrementing $n$. Thus, $n(s, a) \leq J$ for all 
state-action 
pairs $(s, a)$. The estimated reward $\hat{r}$ at interaction $i = J$ is the mean of $\rho$. 
For further interactions $i > J$, the count $n'$ is determined as a function of the 
confidence bound $\phi$ as follows: $n'(s, a)$ is reset to $1$ for all $a \in \A_s$ 
whenever the confidence bound for state-action pair $(s, a)$ decreases below some predefined threshold
$\phi_{min}$ (which ensures a faster return to the default policy). 
Additionally, at interactions $i > J$, the oldest value of the collected reward is dropped, all other 
entries of $\rho$ are shifted one position to the left, and the new collected reward (green cell) is 
inserted at the $J$-th position. Then, 
$\hat{r}$ is computed as the weighted mean of $\rho$ using the corresponding weighting coefficients 
$\gamma_i$. We refer to $\gamma$ as the vector of ``forgetting'' factors (or \textit{discount vector}), 
and it is a tuning parameter along with $\phi_{min}$.

\begin{algorithm*}[tbh]
    \caption{
        \textbf{M}odified \textbf{U}pper \textbf{B}ounding the \textbf{E}xpected Next State 
        \textbf{V}alue (\textbf{UBEV}) Algorithm for \textbf{S}tationary FHMDPs with 
        \textbf{M}oving Window---\textbf{\MUBEV}
        \label{alg:modified_ubev}
    }
    \scriptsize
    \begin{algorithmic}[1]
        \Require
        {
            $\S,\,\A,\,P,\,\Pi_{DP};\,\,\varepsilon, \delta \in \left(0, 1\right];\,\,
            H,\,J,\,M \in \mathbb{N};\,\,r_{max},\phi_{min} \in \mathbb{R};\,\,
            \gamma \in \mathbb{R}^{J};\,\,MWRM \in \{\text{True, False}\}$.
        }
        \State
            $n(s, a) = n'(s, a) = \hat{r}(s, a) = R(s, a) = Q(s, a) = \phi(s, a) = \hat{V}(s, t) = 
            \rho(s, a, j)=0\,\,\forall s \in \S,\,\,a \in \A_s,\,\,t \in [H + 1],\,\,j \in [J]$.
        \State 
            $\delta' = \delta / 9,\,V_{max} = H*r_{max},\,\phi_{+} = 0,\,\pi = \Pi_{DP}$.

        \For{$k=1,2,3...$}{}
            \For{$t=H$\,\,\textbf{to}\,\,$1$}{}\Comment{{\color{blue} Optimistic planning loop}}
                \State
                    $\hat{V}_{t+1} = \hat{V}(\cdot, t+1);\,\,
                    \tilde{V}_{max} = \min\big(\max(\hat{V}_{t+1}), V_{max}\big)$
                \For{$s \in \S$}{}
                    \State
                        $\eta = \ln\big(27*S*A_s*H/\delta'\big)$
                    \For{$a \in \A_s$}{}
                        \State
                            $r = r_{max};\,\,EV = \tilde{V}_{max}$
                        \If{$n'(s, a) > 0$}
                            \State 
                                $\phi(s, a) = \varepsilon*
                                    \sqrt{\frac{2\ln\ln\big(\max\big(e,\,n'(s,\, 
                                    a)\big)\big) + \eta}{n'(s,\,a)}};\,\,
                                \hat{V}_{next} = P(\cdot, s, a) \times \hat{V}_{t+1};\,\,
                                \bar{v} = \max(\hat{V}_{t+1}) - \min(\hat{V}_{t+1})$
                            \State
                                $\bar{v}' = \min\big((H-t)*r_{max}, \bar{v} + \phi_{+}\big);\,\,
                                r = \min\big(r_{max}, \hat{r}(s,a)+ \phi(s, a)\big);\,\,
                                EV = \min\left(\tilde{V}_{max},\hat{V}_{next} + \bar{v}'*\phi(s, a)\right)$
                        \EndIf
                        \State $Q(s, a) = r + EV$
                    \EndFor 

                    \State
                        $Q_{max} = \max_{a\in\A_s}\big(Q(s,a)\big)$
                    \If {$\,\,Q_{max}$ duplicates in $Q(s,\cdot)\,\,$}
                        \If {$\,\,Q_{max} = Q\big(s, \Pi_{DP}(s, t)\big)\,\,$}
                            $\,\,\tilde{a} = \Pi_{DP}(s, t)\,\,$
                        \textbf{else}
                            $\,\,\tilde{a} \sim \mathcal{U}\big(\{a\,\vert\,Q(s, a) = Q_{max}\}\big)$
                        \EndIf
                    \Else
                        $\,\,\tilde{a} = \argmax_{a\in\A_s}Q(s, a)$
                    \EndIf
                    \State
                        $\pi_{k}(s, t) = \tilde{a};\,\,\hat{V}(s, t) = Q(s, \tilde{a})$

                    \If{$\,\,n'(s, \tilde{a}) > 0\,\,$}
                        $\,\,\tilde{\phi} = \varepsilon*\sqrt{\frac{2\ln\ln\big(\max\big(e,\,
                        n'(s,\,\tilde{a})\big)\big) + \eta}{n'(s,\,\tilde{a})}};\,\,
                        \phi_{+} = \max\big(4*\sqrt{S}*H^2*\tilde{\phi}, \phi_{+}\big)$
                    \EndIf
                \EndFor 
            \EndFor 

            \State
                $\tilde{s} = \big[s_1^{(1)}, ..., s_1^{(M)}] \sim \mathcal{U}\big(1, S\big),\,
                s_1^{(i)} \neq s_1^{(j)}\,\,\,\forall i, j \in [1, M]$
            \For{$m=1$\,\,\textbf{to}\,\,$M$,\,\,$t=1$\,\,\textbf{to}\,\,$H$}{}
            \Comment{{\color{blue} Execute policy for one episode}}
                \State
                    $a_t^{(m)} = \pi_k(s_t^{(m)}, t);\,\,\bar{a}_t^{(m)} = \Pi_{DP}(s_t^{(m)}, t);\,\,
                    s_{t+1}^{(m)} \sim P(\cdot \vert s_t^{(m)}, a_t^{(m)});\,\,
                    \bar{s}_{t+1}^{(m)} \sim P(\cdot \vert s_t^{(m)}, \bar{a}_t^{(m)})$
                \State
                    $r^{(m)}_t = \R\big(s_t^{(m)}, s_{t+1}^{(m)}, \bar{s}_{t+1}^{(m)});\,\,
                    R\big(s_t^{(m)}, a_t^{(m)}\big) += r^{(m)}_t$
                    \Comment{{\color{blue} Call to Function~\hyperref[alg:reward_func]{1},
                    accumulate rewards}}
                    \If{$MWRM = \text{True}$}\Comment{{\color{blue} Employ moving window}}
                    \State
                        $j_m \coloneqq n\big(s_t^{(m)}, a_t^{(m)}\big);\,\,
                        j_m' \coloneqq n'\big(s_t^{(m)}, a_t^{(m)}\big)$
                    \If{$j_m \leq J$}
                        \State
                            $\rho\big(s_t^{(m)}, a_t^{(m)}, j_m\big) = r^{(m)}_t;\,
                            n\big(s_t^{(m)}, a_t^{(m)}\big)\tiny{++};\,
                            n'\big(s_t^{(m)}, a_t^{(m)}\big)\tiny{++};\,
                            \hat{r}\big(s_t^{(m)}, a_t^{(m)}\big) = R\big(s_t^{(m)}, a_t^{(m)}\big) \big/ 
                                n\big(s_t^{(m)}, a_t^{(m)}\big)$
                    \Else
                        \State
                            $R\big(s_t^{(m)}, a_t^{(m)}\big) = 
                            R\big(s_t^{(m)}, a_t^{(m)}\big) - \rho\big(s_t^{(m)}, a_t^{(m)}, 1\big)$
                            \Comment{{\color{blue}``Forget'' the oldest reward}}
                        \State
                            $\rho\big(s_t^{(m)}, a_t^{(m)}, \cdot\big) = 
                            \rho\big(s_t^{(m)}, a_t^{(m)}, \cdot\big)_{2:J}
                            \doubleplus 
                            \big\{r^{(m)}_t\big\}$
                            \Comment{{\color{blue}Incorporate the newest reward}}
                        \State
                            $\hat{r}\big(s_t^{(m)}, a_t^{(m)}\big) = 
                            \Big(\gamma \times \rho\big(s_t^{(m)}, a_t^{(m)}\big)\Big) \big/J$
                        \If{$\,\,\phi\big(s_t^{(m)}, a_t^{(m)}\big) \leq \phi_{min}\,\,$}
                            $\,\,n'\big(s_t^{(m)}, a\big) = 1\,\,\forall a \in \A_s\,\,$
                        \textbf{else}
                            $\,\,n'\big(s_t^{(m)}, a_t^{(m)}\big)++$
                        \EndIf
                    \EndIf
                \Else\Comment{{\color{blue} Do not apply moving window}}
                    \State
                        $n\big(s_t^{(m)}, a_t^{(m)}\big)++;\,\,n'\big(s_t^{(m)}, a_t^{(m)}\big)++;\,\,
                        \hat{r}\big(s_t^{(m)}, a_t^{(m)}\big) = R\big(s_t^{(m)}, a_t^{(m)}\big) \big/ 
                            n\big(s_t^{(m)}, a_t^{(m)}\big)$
                \EndIf
            \EndFor 
        \EndFor 
    \end{algorithmic}
\end{algorithm*}

The notation for \MUBEV\ (Algorithm~\ref{alg:modified_ubev}) is as follows. $\S$ is the set of states;
$\A_s$ is the set of allowable actions in state $s$, and thus, 
$\A = \bigcup_{s \in \S} \A_s$ is the total action set; 
$S$, $A_s$ and $A$ denote cardinality of finite sets $\S$, $A_s$ and $\A$, respectively; 
$P$ is a 3-dimensional tensor of predefined transition probabilities;
$\Pi_{DP}$ is the default path (DP) policy, computed using an SP algorithm as the initial estimate for 
$\pi$;
$\varepsilon$ is the exploration bonus, and $\delta$  is the failure probability (see~\cite{ubev} for details);
$H$ is the length of the MDP's time horizon; 
$J$ is the size of moving window;
$M$ is the number of \MUBEV\ tokens;
$r_{max}$ is the maximum reward an agent can receive per transition;
$\phi_{min}$ is the minimum value of the confidence bound which determines when to reset the collected 
statistics (i.e. counts $n'$);
$\gamma$ is a vector of discounts for the rewards stored in the reward memory $\rho$;
$MWRM$ is a boolean flag to enable/disable MWRM;
$n(s, a)$ and $n'(s, a)$ count the times action $a$ is played in state $s$ (see previous paragraph for
details);
$\hat{r}(s, a)$ and $R(s, a)$ are the normalised and accumulated rewards in state $s$ under action 
$a$, respectively;
$Q(s, a)$ is the Q-function~\cite{ubev} for state $s$ and action $a$;
$\phi(s, a)$ is the confidence bound for state $s$ and action $a$;
$\hat{V}(s, t)$ is the value function from time step $t$ for state $s$;
$\rho(s, a, j)$ represents a reward stored in a 3-dimensional tensor of reward memory $\rho$ 
for state $s$ and action $a$ at memory epoch $j$;
$\delta'$ represents a scaled failure tolerance (see~\cite{ubev} for details);
$V_{max}$ is the maximum value of the value function for next states;
$\phi_{+}$ is the correction term which keeps track of the largest confidence bound for the least visited
state-action pair under the policy $\pi$~\cite{pmlr-v80-zanette18a};
$\hat{V}(\cdot, t+1)$ and $P(\cdot, s, a)$ denote vectors of length $S$, and
$Q(s, \cdot)$ is interpreted as a vector of length $A_s$;
$e$ is the Euler's number; 
$\eta$, $r$, $EV$, $\bar{v}$ and $\bar{v}'$ are auxiliary variables;
$\bar{v}$ represents the range of vector $\hat{V}_{t+1}$~\cite{pmlr-v80-zanette18a};
vector $\tilde{s}$ is a vector of initial states of \MUBEV\ tokens, which is uniformly sampled 
in range from $1$ to $S$ with no repeated entries; $m$ is the index of an agent (a vehicle with a token)
that interacts with the environment each time step $t$, and receives reward $r_t^{(m)}$ determined by 
the reward function defined in Function~\hyperref[alg:reward_func]{1}.


%% file: 03_04_reward_func.tex

The reward function (Function~\hyperref[alg:reward_func]{1}) returns the total reward $r_t$ at time $t$, which implicitly incorporates
two objectives, namely travel distance and parking availability.

\begin{algorithm}[h]
    \scriptsize
        \begin{algorithmic}[1]
            \Require{$\alpha \in \left[\alpha_{min}, 1\right]; s_t, s_{t+1}, \bar{s}_{t+1} \in \S;
                \beta_1, \beta_2, r_{max} \in \mathbb{R}$.}
            \Ensure{$r_t$.}
            \State
            $s_t \coloneqq s, s' \coloneqq s_{t+1}, \bar{s}' \coloneqq \bar{s}_{t+1}$
            \Function{$\R$}{$s, s', \bar{s}'$}
                \State
                Get the state lengths $\lambda_{s'}$ and $\lambda_{\bar{s}'}$, 
                and the total roadside parking capacity $c_{\bar{s}'}$.
                \State
                Observe the current roadside parking availability $\nu_{s'}$.
                \State
                $w_{\bar{s}'} = \mathcal{W}_G(\alpha, \bar{s}', \lambda_{\bar{s}'}, c_{\bar{s}'}),\,\,$
                $w_{s'} = \mathcal{W}_G(\alpha, s', \lambda_{s'}, \nu_{s'})$
                \Comment{{\color{blue}Compute weights for $\bar{s}', s'$, respectively, using 
                Equation~\ref{alg:weight_func}}}
                \If {$w_{\bar{s}'} \neq w_{s'}$}\Comment{{\color{blue} Uncertainty has been encountered}}
                    \State
                        Get route total weights $\omega_{s'}$ and $\omega_{\bar{s}'}$.
                        $\Omega_{s, \bar{s}'} = \omega_{\bar{s}'} + w_{\bar{s}'};\,\,
                        \Omega_{s, s'} = \omega_{s'} + w_{s'};\,\,
                        r_t = 1 - \frac{\Omega_{s, s'}}{\Omega_{s, \bar{s}'}}$
                    \If {$\,s' = \bar{s}'\,$}
                        $\,r_t = \beta_1 * r_t$
                        \Comment{{\color{blue} Agent takes the DP action}}
                    \Else
                        $\,\,\,r_t = \beta_2 * r_t$
                        \Comment{{\color{blue} Agent takes an alternative action}}
                    \EndIf
                \Else
                    $\,\,\,r_t = r_{max}$
                    \Comment{{\color{blue} There is no uncertainty}}
                \EndIf
                \Return $r_t$
            \EndFunction
    \end{algorithmic}
    \caption*{\textbf{Function 1} The Reward Function\label{alg:reward_func}}
\end{algorithm}

Function~\hyperref[alg:reward_func]{1} receives three arguments: (i) $s_t$, the previous state of an agent;
(ii) $s_{t+1}$, the current state;
and (iii) $\bar{s}_{t+1}$, the state to which the agent would ``jump''
if the default action in state $s_t$ 
were taken. To begin with, $\lambda_{s'}$ and $\lambda_{\bar{s}'}$ are the lengths 
of road links included in states $s'$ and $\bar{s}'$, respectively, $c_{\bar{s}'}$ is the total 
on-street parking capacity for state $\bar{s}'$, and $\nu_{s'}$ is the current parking availability in 
state $s'$. Note that the state length $\lambda$ can be easily computed for a given road network. Also,
as we assume that total roadside PS capacity for the road network is available (e.g. via digital map provider),
and thus the total state capacity $c$ is known for each state. However, PSA is subject to real-time measurements
(e.g. obtained via \cite{wu2019early, bock2015street}).
$w_{\bar{s}'}$ is the state weight if the agent would transit from state $s$ 
to state $\bar{s}'$ in which all the on-street parking stops are treated as available. In contrast, 
$w_{s'}$ represents the state weight of the actual transition from state $s$ to 
state $s'$ at time $t$.
If uncertainties occur in the environment, then the weights computed in line 7 
are not equal to each other, and thus the DP policy has to be amended. In this case, 
the reward is computed using the total route weight of two (in general different) routes. Further, $\Omega_{s, \bar{s}'}$ and $\Omega_{s, s'}$ are the total weights of routes 
from $s$ to the destination state $s_f$ via $\bar{s}'$ and $s'$, respectively, and
$\omega_{\bar{s}'}$ and $\omega_{s'}$ are the total weights of routes from $\bar{s}'$ and $s'$ to the destination $s_f$, respectively (see Figure~\ref{fig:moving_window}-Right).
As a result, $\Omega_{s, \bar{s}'} = \omega_s$, and since $\Omega_{s, \bar{s}'}<\Omega_{s, s'}$,
reward computed in line 7 always satisfies $r_t < 0$.
Finally, $\beta_1$ and $\beta_2$ are 
used to speed up the learning process of searching for a detour and returning to the DP route
respectively, and their values may depend on many factors including values of $\alpha$
and $r_{max}$ or the road network itself. To achieve a better and faster 
performance of the algorithm, $\beta_1$ and $\beta_2$ need to be increasing functions of $\alpha$ as
illustrated in the following section. 
Finally, if a given state is not affected by uncertainties, the 
agents receive the maximum reward $r_{max}$ available per transition.


%% file: 04_evaluations.tex

In this section we are interested in evaluating a route recommender system for 
smart parking, for which a number of \MUBEV\ tokens is distributed to emulate real vehicles 
probing 
the uncertain environment. These tokens are passed from vehicle to vehicle using the same DLT framework 
introduced in~\cite{sptoken_journal, iccve}, where vehicles in possession of tokens are permitted to 
write data to the 
DLT. The token passing mechanism is 
dictated by both the \MUBEV\ algorithm and the DLT architecture and can be implemented using a MEC/cloud-based 
service (also described in~\cite{sptoken_journal, iccve}). 
Once the unknown environment has been ascertained, route recommendations can be accessed via a 
smart parking app by a variety of users interested in optimal routing.

\begin{figure}[h]
    \centering
    \includegraphics[width=0.5\columnwidth]{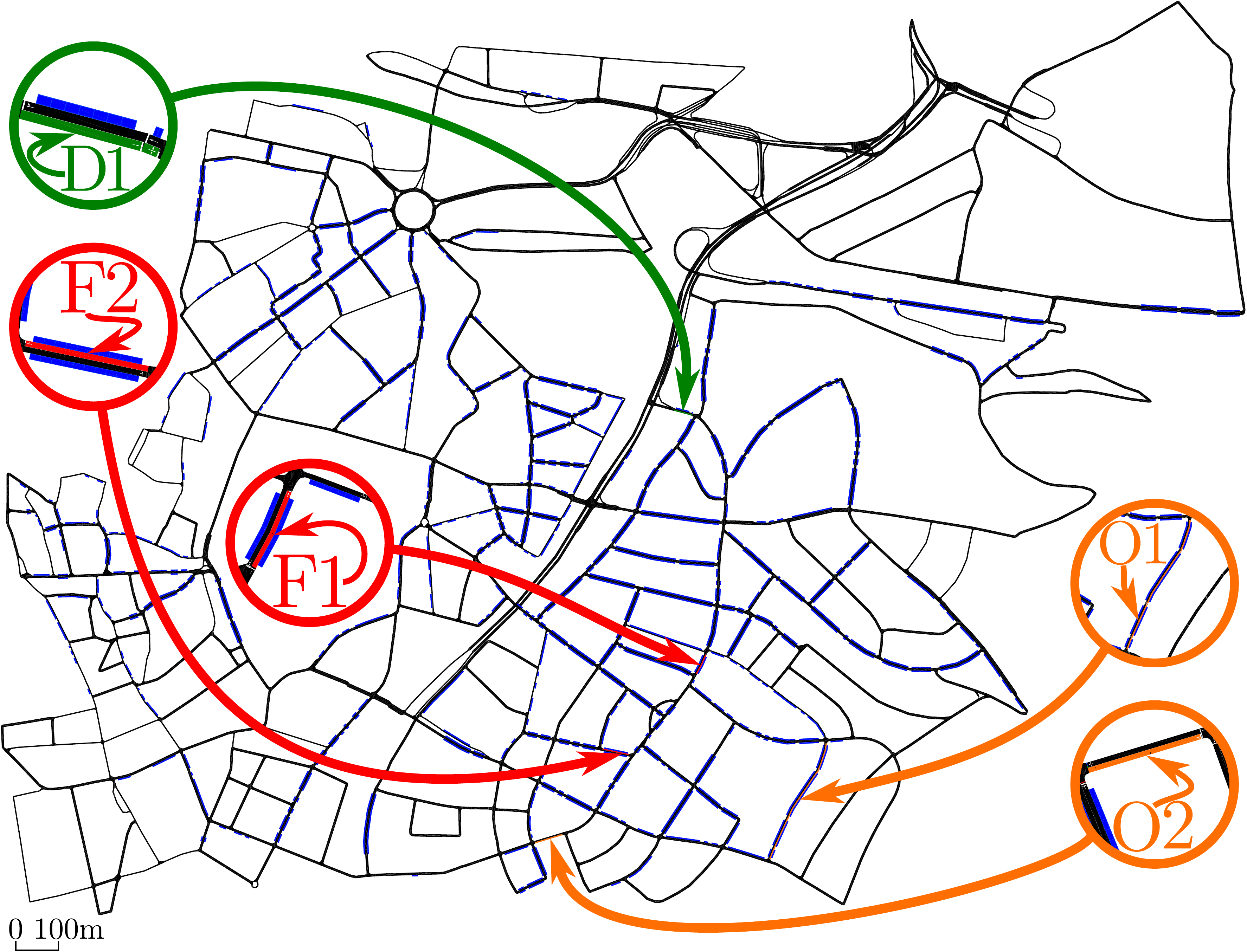}
    \caption{The road network used in the experiments comes from an urban area close to the city centre in W\"urzburg, Germany.
        In the OpenStreetMap network of this area\protect\footnotemark, PSs have been tagged, and thus
        on-street PSs can be easily imported to a SUMO road network file. Some road segments of interest are highlighted, 
        representing either origins (O), destinations (D), or sources of full parking areas (F).}
    \label{fig:road_network}
\end{figure}
\footnotetext{\scriptsize{\url{https://www.openstreetmap.org/\#map=14/49.7830/9.9553&layers=T}}}

To evaluate \MUBEV, a number of numerical experiments was designed 
using the traffic simulator SUMO \cite{sumo}. Interactions with running simulations (including those to measure roadside parking occupancy) are accomplished using the Python programming language and SUMO packages \textit{Sumolib} and \textit{TraCI}. The generic 
settings for our simulations are as follows:
\begin{itemize}
    \item The urban area in W\"urzburg, Germany shown in Figure~\ref{fig:road_network} is used for traffic simulations in all our 
        experiments. Once all the U-turns are removed\footnote{U-turns may lead to undesirable recurrent attempts to use DP policy.}
        from the network file, the road network results in a graph with 861 nodes (states).
    \item A number of road links are selected as origins, destinations, and sources of full on-street 
        parking areas (see Figure~\ref{fig:road_network} for details). 
        Specifically, we use: origin $O1$ and full parking area (Full-PA) $F1$ in Experiment~\hyperref[sec:exp1]{1} 
        and~\hyperref[sec:exp2]{2}; origin $O2$ and Full-PA $F2$ in Experiment~\hyperref[sec:exp3]{3};
        destination $D1$ in all our experiments. 
    \item In all our simulations, we create and use a vehicle type based on the default SUMO vehicle 
        type with maximum speed 118.8 km/h and 
        impatience\footnote{\scriptsize{\url{https://sumo.dlr.de/wiki/Definition_of_Vehicles_Vehicle_Types_and_Routes}}.} 
        equal to 0.5. When these vehicles are in possession of a token, they become 
        \textit{virtual} \MUBEV\ \textit{vehicles}. 
        In Experiment~\hyperref[sec:exp1]{1} and~\hyperref[sec:exp2]{2}, these vehicles have 
        no parking stops assigned. In Experiment~\hyperref[sec:exp3]{3},
        every 100th vehicle added to the simulation is assigned to a random parking stop along its route. 
        For the generation of Full-PAs, we release a number of cars of the 
        aforementioned vehicle type and populate the selected on-street parking lanes with them.
    \item For all our experiments, we add an average of 250 background cars with random routes, which can 
        carry tokens if required. 
    \item DPs are calculated using Dijkstra's SP algorithm (SciPy\footnote{\scriptsize{\url{https://docs.scipy.org/doc/scipy/reference/generated/scipy.sparse.csgraph.dijkstra.html}}} implementation).
    \item A state of an RL agent is considered to be a set of road links joined into one state using 
        the merging technique introduced in~\cite{sptoken_journal}. We refer to a token trip as an RL
        episode. Before policy execution, the tokens are uniformly distributed on the road network.
        The length of the MDP's time horizon $H$ is set to 50. $H$ determines a maximum number of 
        allowed road links that each token can traverse over 
        a given episode. If tokens do not reach a specified destination within this restriction, their 
        trips will be declared incomplete (i.e. unsuccessful). 
\end{itemize}
\hspace{0.5em}

Regarding the input parameters of Algorithm~\ref{alg:modified_ubev} and 
Function~\hyperref[alg:reward_func]{1}, in all our experiments 
we set $\varepsilon = \delta = r_{max} = 1$, $J = 10$, $\phi_{min} = 1.25$.
The components of vector $\gamma$ are computed as follows: 
$\gamma_j = 1 - (J - j)*(1 - \gamma_{min})/(J - 1)$, $j \in [J]$, with $\gamma_{min}$ empirically 
set to $0.1$. 
In Experiment~\hyperref[sec:exp1]{1}, we set $\beta_1 = \beta_2 = 20$ for simulations with 
$\alpha = 0.66$, and $\beta_1 = \beta_2 = 60$ for $\alpha = 0.80$.
In Experiment~\hyperref[sec:exp2]{2} and~\hyperref[sec:exp3]{3}, we set
$\beta_1 = \beta_2 = 20$ and $\alpha = \alpha_{min} = 0.66$.
Any specific additional setting for each individual experiment will be described in the corresponding 
subsection below.

\subsection{\myfontsize{11}{\textnormal{\textit{Experiment 1: Optimal route estimation under 
    uncertainty using different values of $\alpha$}}}
    \label{sec:exp1}
    }

The objectives of this experiment is to illustrate that our DLT-enabled RL approach is able to 
determine a simple uncertain environment using a single token for different values of $\alpha$. 
For a chosen OD pair, we artificially simulate a Full-PA on some road links 
along DP at various time intervals. In this scenario, our goal is to show that the 
token-enabled \MUBEV\ algorithm can distinguish between Free Parking Intervals (Free-PIs) and Full Parking Intervals (Full-PIs), 
and, in the latter case, find the next optimal route for the selected OD pair.

\begin{figure}[h]
    \centering
    \subfloat{{\includegraphics[width=0.45\columnwidth]{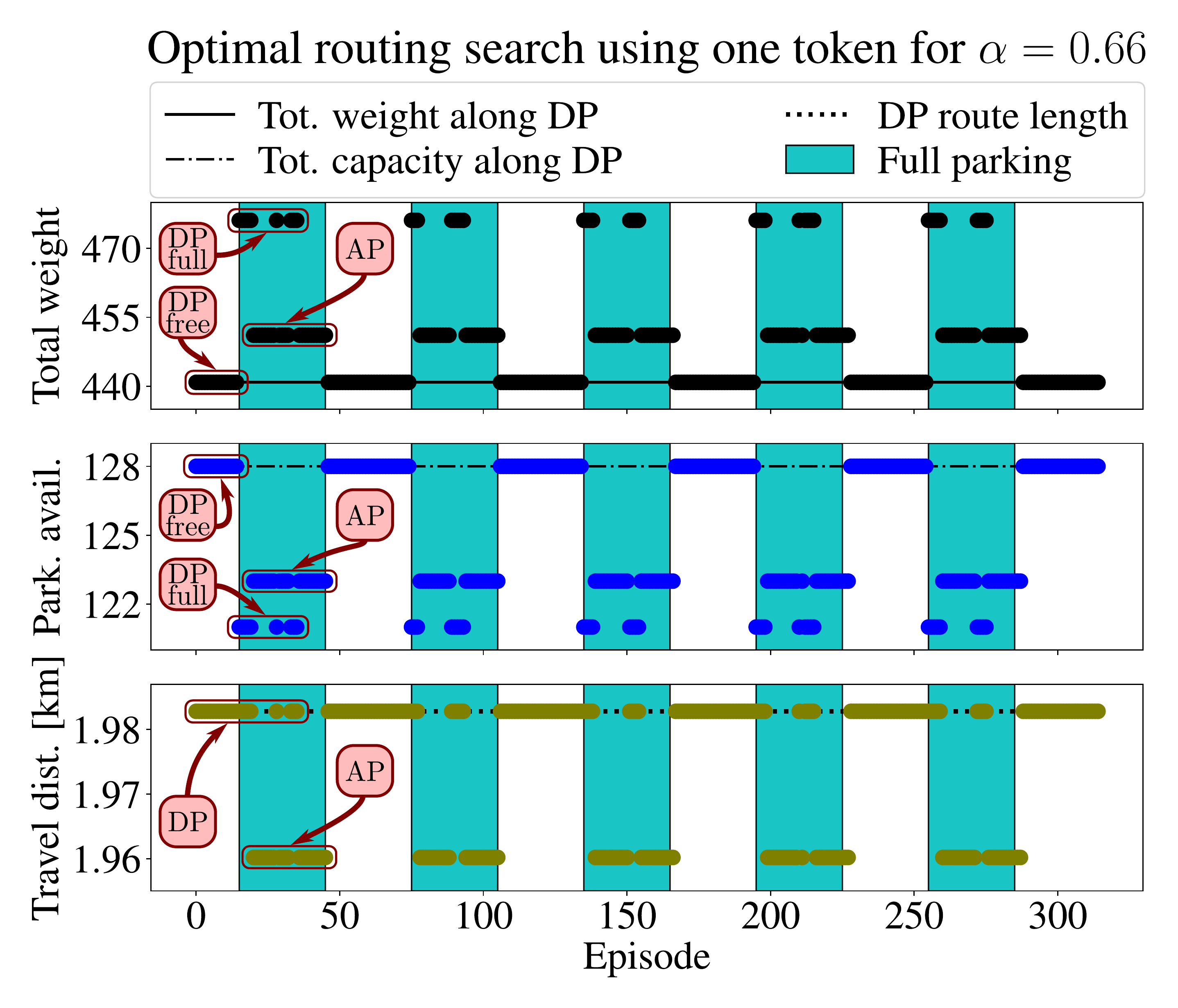}}}
    \subfloat{{\includegraphics[width=0.45\columnwidth]{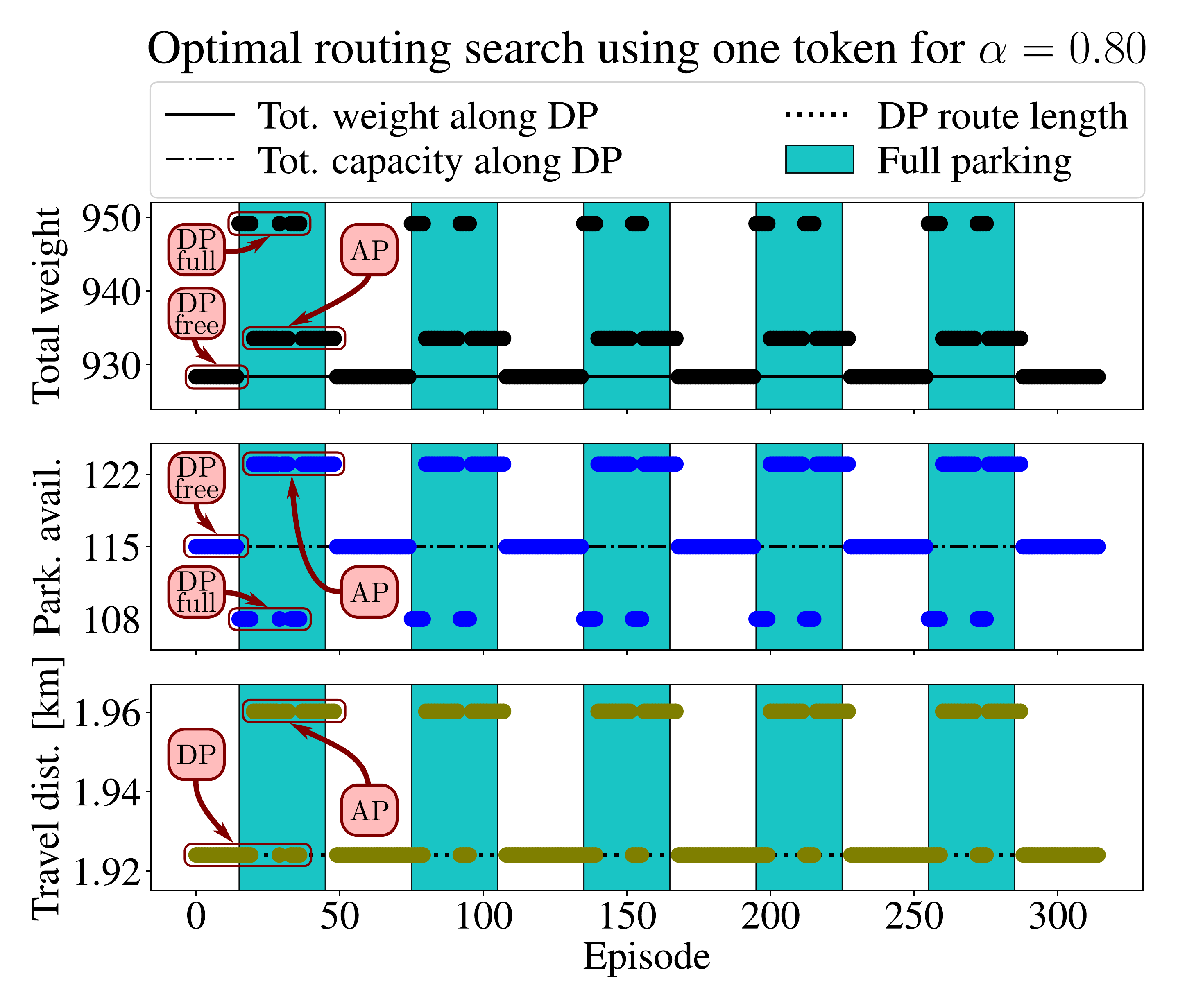}}}
    \caption[na]{
        Experiment 1: Performance indices of \MUBEV\ with MWRM
        during the learning process on a changing environment using one token and fixed OD pair $\{O1, D1\}$
        for $\alpha=0.66$ (\underline{Left}) and $\alpha=0.80$ (\underline{Right}).
        Annotations are only provided for the first part of the learning process (DP: default path, AP:
        alternative path).
    }
    \label{fig:exp1}
\end{figure}

Particularly, this illustrative experiment is designed as follows. A single \MUBEV\ token is used over
each episode of the learning process to collect data used to update the MDP's policy. 
This token has a fixed OD pair, namely $\{O1, D1\}$ as marked in Figure~\ref{fig:road_network}. 
Additionally, we choose $F1$ (a set of road segments which belong to DP for the 
specified OD pair) and generate a Full-PA in it at different time intervals. Over each 
episode, we start the token from $O1$ and ask it to travel to $D1$, keeping a record of its performance 
in terms of route characteristics, such as the total weight (computed using
Equation~\ref{alg:weight_func}), 
the number of available on-street PSs, and the travel distance (route 
length), regardless of its success in attempting to reach $D1$. During Full-PIs, 
the total weight of the token route enlarges while parking availability decreases 
(this is annotated with ``DP full'' for the first Full-PI in Figure~\ref{fig:exp1}).
\MUBEV\ is expected to provide an alternative path (AP) which is an optimal solution leading to a 
minimum possible value of total weight 
during Full-PIs (as also annotated in Figure~\ref{fig:exp1}),
and that DP routing would eventually be advised in all other situations.

The results of a single realisation of the previously described experiment are depicted in 
Figure~\ref{fig:exp1} for different values of $\alpha$. As it can be noted, the token succeeds 
in avoiding Full-PAs once they are created
(which can be better observed in the travel distance plots),
and learns an AP route with a smaller total weight 
than that of DP during Full-PIs
(annotated as ``DP full''). 
The token also successfully returns to DP once Full-PAs 
are cleared. Both actions, avoiding and returning, happen within a reasonably small number of episodes.
Additionally, thanks to MWRM, this rapid adaptation is present even at the beginning of the learning 
process, and the performance of \MUBEV\ is nearly uniform as time passes. These observations 
validate our expectations about \MUBEV\ concerning its ability to adapt rapidly to uncertain environments.

When comparing the right and left panels of Figure~\ref{fig:exp1}, we can notice that the length of DP, 
and the total parking capacity along it, are larger for a smaller value of $\alpha$ (left panel) than 
the corresponding values for a larger $\alpha$ (right panel). This indicates, as expected, that 
smaller values of $\alpha$ generally lead to longer DP routes. Consequently, it would be impractical 
to consider an extremely small value of $\alpha$ (even if
it satisfies $\alpha \in [\alpha_{min}, 1]$), as it could lead to a dramatic increase in travel 
distances/times. As it can also be seen in Figure~\ref{fig:exp1}, our \MUBEV\ performs as expected for 
both values of $\alpha$.

\subsection{\myfontsize{11}{\textnormal{\textit{Experiment 2: Comparative analysis of \MUBEV\ 
    with and without MWRM}}}
    \label{sec:exp2}
}

\begin{figure}[h]
    \centering
    \subfloat{{\includegraphics[width=0.45\columnwidth]{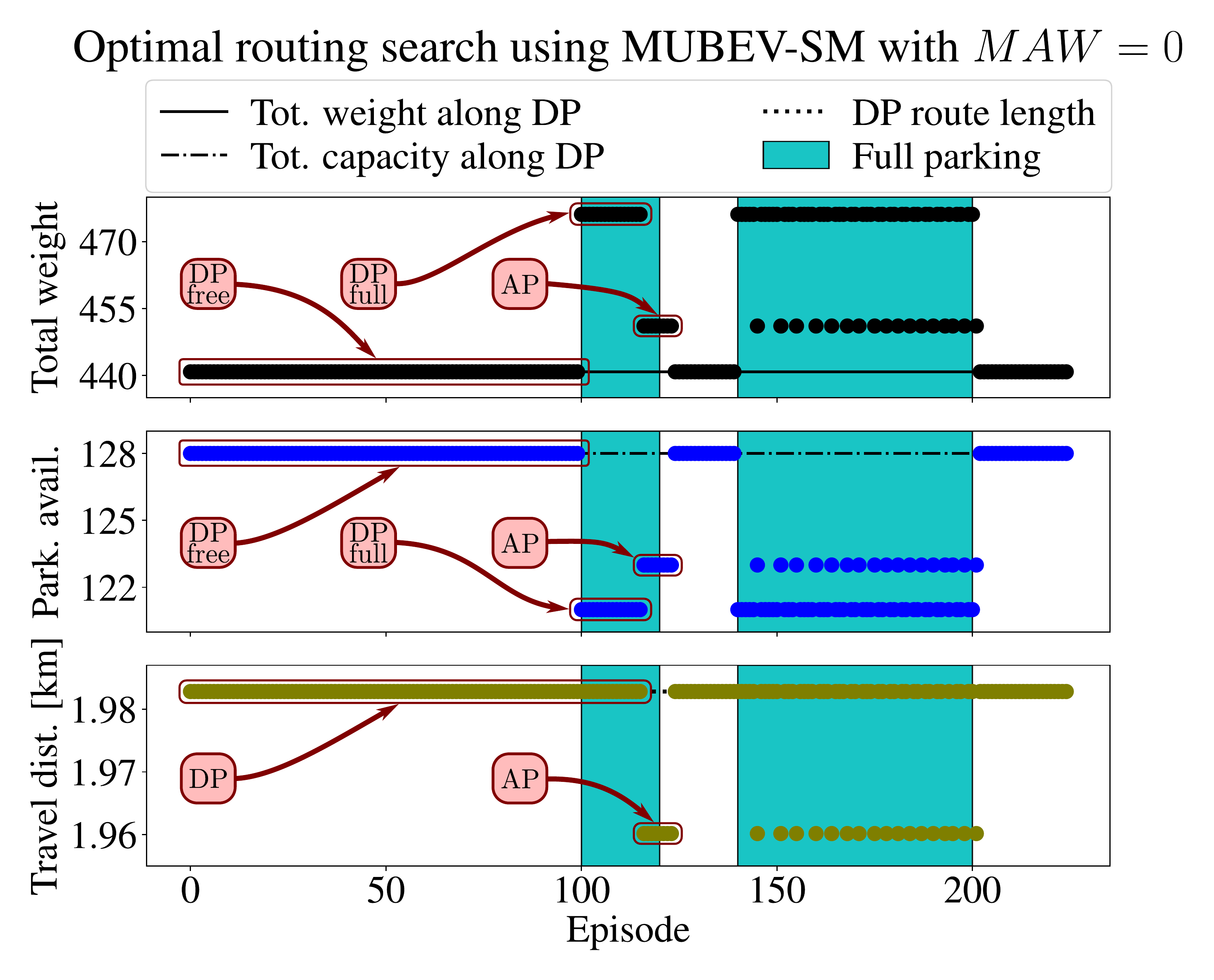}}}
    \subfloat{{\includegraphics[width=0.45\columnwidth]{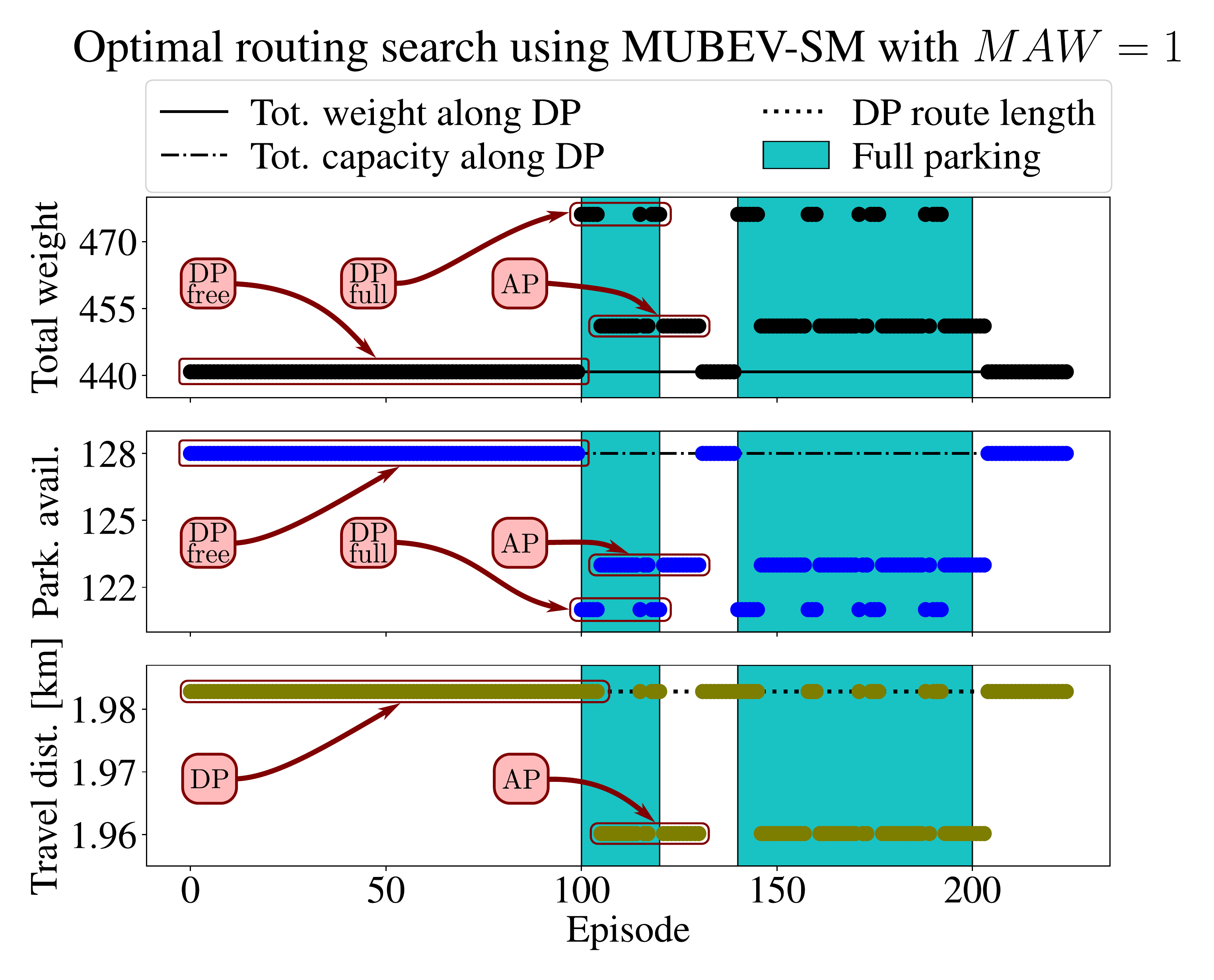}}}
    \caption{Experiment 2: Comparative performance of \MUBEV\ without (\underline{Left}) and 
        with (\underline{Right}) MWRM on a changing environment with 
        non-uniform Full-PIs, using a single token and fixed OD pair.
        Annotations are only provided for the first part of the learning process (DP: default path, AP:
        alternative path).
    }
    \label{fig:exp_maw}
\end{figure}

In this second experiment, we want to evaluate the performance of \MUBEV\ with and 
without MWRM using a single token and non-uniform Full-PIs. For this, we use a 
similar setup as in Experiment~\hyperref[sec:exp1]{1} (i.e. same OD pair, Full-PA on 
$F1$, and one token probing the environment). To begin with, the token starts 
at $O1$ and is asked to travel to $D1$ each episode, keeping a record of the performance indices as
mentioned in the previous subsection. We intentionally do not introduce a Full-PA for a long 
period of time, allowing the token to collect lots of ``positive'' rewards while traveling along DP,
after which we generate a Full-PA over a reasonable short time interval. Afterwards, we allow a 
short Free-PI followed by a Full-PA on $F1$ again, but over a significantly longer period of time. 
The results of this comparative analysis for a single realisation are shown in Figure~\ref{fig:exp_maw}.

From Figure~\ref{fig:exp_maw}-Left, we can notice that the learning process using \MUBEV\ without 
MWRM is notably slow during the first Full-PI  
(better observed in the travel distance plot, left panel),
meaning that a large number of episodes are required to learn an AP.
On the contrary, the algorithm is able to provide an optimal 
detour AP much faster (in terms of fewer episodes) if MWRM is applied as seen in 
Figure~\ref{fig:exp_maw}-Right. 
The latter configuration also allows for a better performance during the second Full-PI: \MUBEV\
is able to learn an AP route in a
small number of episodes, and such an AP solution is preferred over the Full-PI. It must be noted that without 
MWRM, the system mostly selects DP routing over the whole learning process, which can be seen in the lowest subplot 
in Figure~\ref{fig:exp_maw}-Left. The reason of such performance is that the system without 
MWRM has collected lots of ``positive'' rewards and statistics during the first (long) Free-PI.

Finally, it can be observed that returning to DP (once the Full-PAs are cleared)
takes slightly longer with MWRM than without. 
However, with MWRM, the system does not require much tuning of its parameters for each particular 
uncertainty, and performs significantly better in general.
Note that experiments with a single token are useful to analyse the performance of \MUBEV\
in uncertain environments, but more elaborated experiments must be performed for a comprehensive validation. Experiment~\hyperref[sec:exp3]{3} addresses this.

\subsection{\myfontsize{11}{\textnormal{\textit{Experiment 3: Route recommendations from a \MUBEV-based 
    system, and speedup in learning}}}
    \label{sec:exp3}
}

\begin{figure}[h]
    \centering
    \subfloat{{\includegraphics[width=0.45\columnwidth]{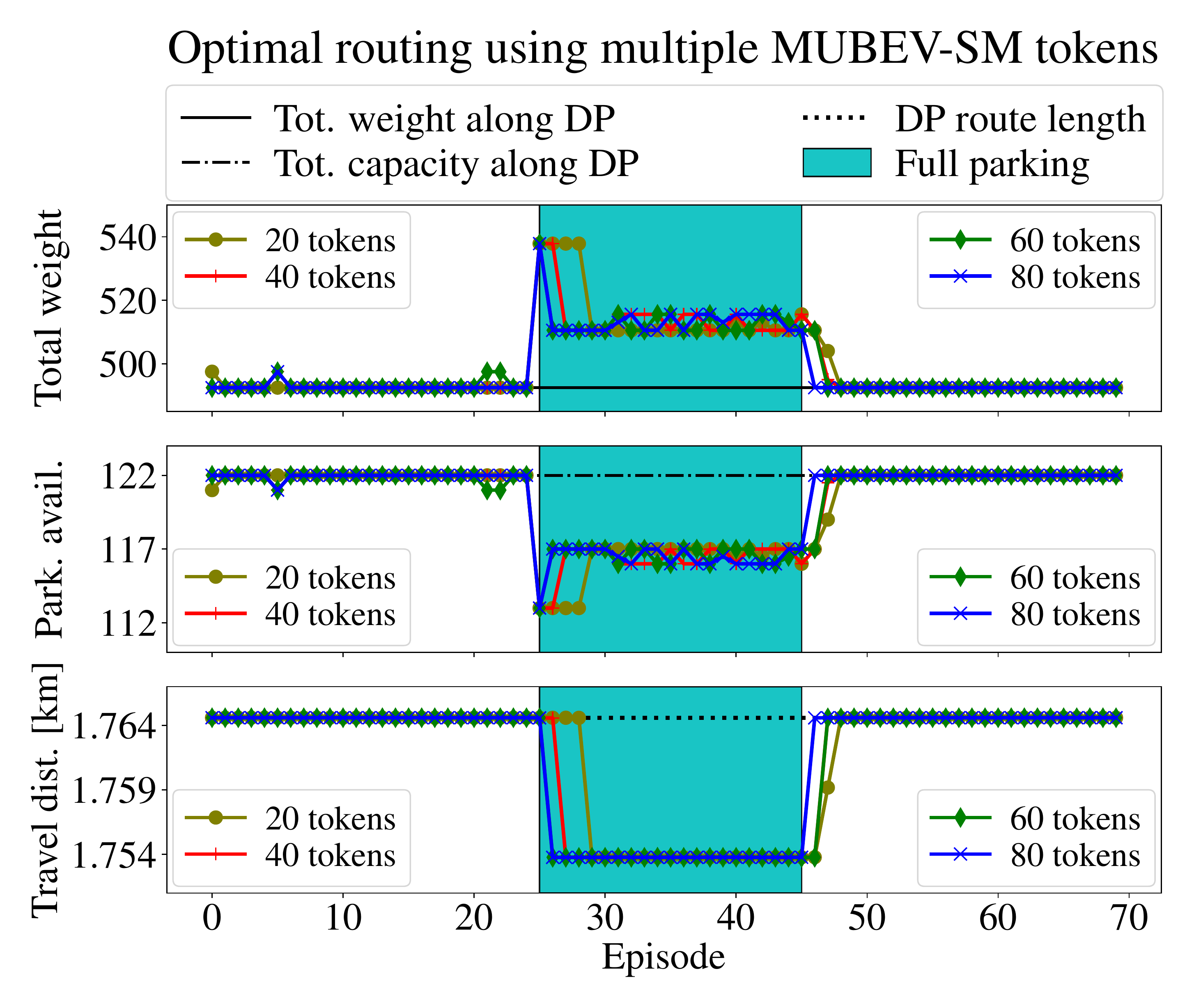}}}
    \subfloat{{\includegraphics[width=0.45\columnwidth]{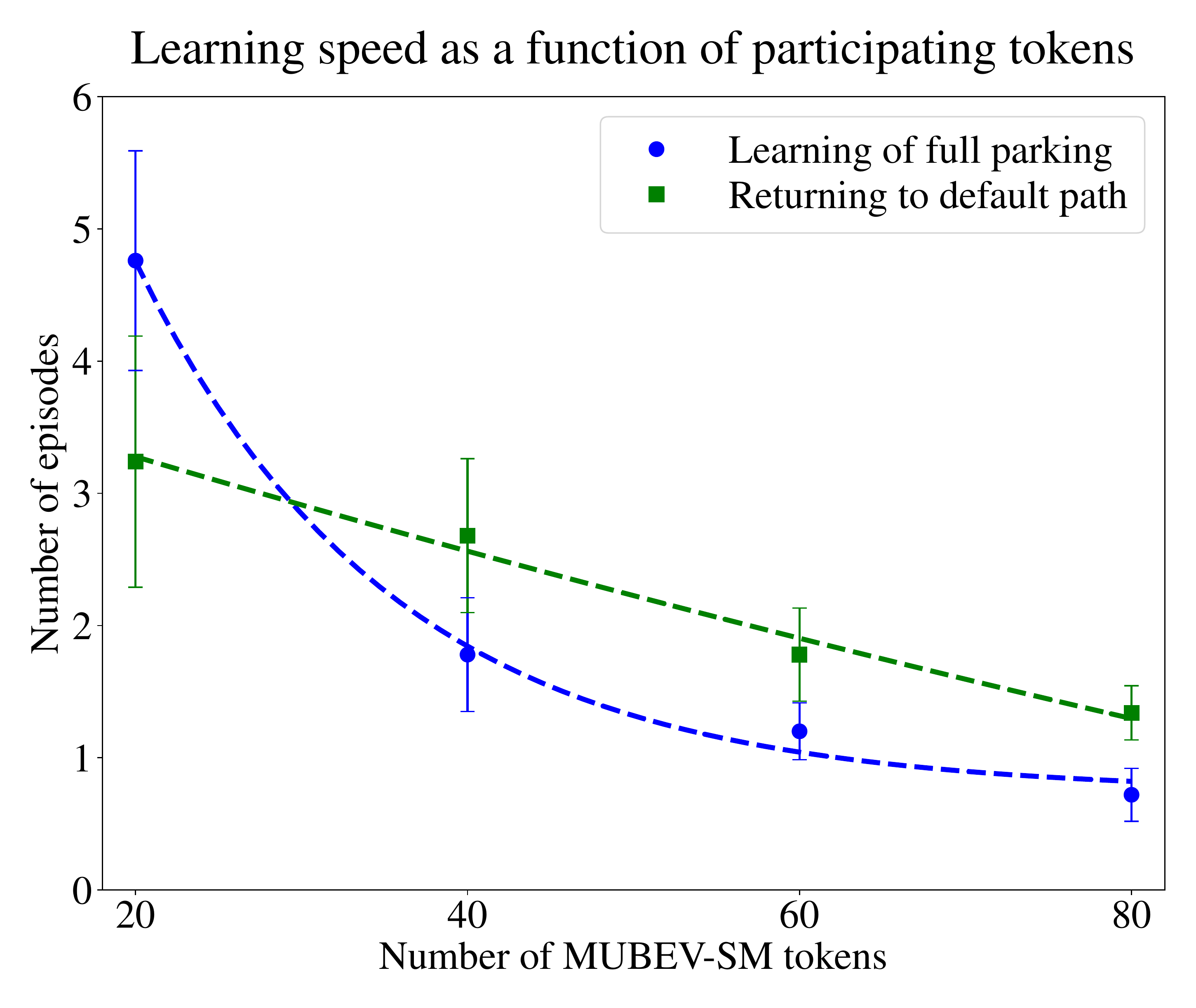}}}
    \caption{Experiment 3: 
        \underline{Left:} Median values of route attributes for a test vehicle using 
        recommendations from a \MUBEV-based routing system involving multiple tokens. 
        Each datapoint represents the median value of 50 different realisations of the experiment.
        \underline{Right:} Average learning speed using multiple tokens. The error bars depict the 
        95\% confidence interval for the mean. Dashed lines were obtained using exponential curve fitting.
    }
    \label{fig:exp_mult}
\end{figure}

Along with an illustration of how \MUBEV\ route recommendations can be delivered to a number of drivers,
this experiment is also designed to evaluate the performance of the system as a function of the number of 
tokens used to collect the data to update the MDP's policy. The setup for this experiment is as 
follows. Over each episode, we release a set of tokens 
starting at different origins according to a uniform spatial distribution. All these tokens have 
a common destination $D1$. In parallel, we analyse the performance of a \textit{test} 
(non-\MUBEV) \textit{vehicle} starting at origin $O2$ every single episode. The test
vehicle is asked to travel to destination $D1$ using recommendations provided by a \MUBEV-based 
routing system, whose recommendations are defined as follows: (i) a route connecting $\{O2, D1\}$ is calculated
based on the estimated MDP's policy; and (ii) whenever a complete route for $\{O2, D1\}$ 
cannot be calculated using the \MUBEV\ policy (e.g. the destination is not reached within the length 
horizon $H$), the latest valid recommendation is reused.
A Full-PA is generated on the set of road links $F2$ (which belong to the
DP for $\{O2, D1\}$) during episodes 25 to 45. It is worth noting that a new test car 
is released at the end of each episode as soon as the policy is updated. Figure~\ref{fig:exp_mult} depicts 
the results for this experiment obtained from 50 different realisations.

As it can be concluded from Figure~\ref{fig:exp_mult}-Left, the recommender system has a 
remarkable performance for different number of participating tokens. However, 
the number of tokens directly affects the convergence rate of the \MUBEV\
algorithm: the more tokens involved, the faster the learning process. A more conclusive relationship between 
the number of participating tokens and the average number of episodes required to learn a new 
traffic condition (either free or full parking areas) is depicted in Figure~\ref{fig:exp_mult}-Right.
This relationship reflects an exponential-like decay for learning of Full-PAs, and a nearly linear 
decrease for returning to DP.

%% file: 05_conclusions.tex

We propose a routing system design based on DLT and multi-agent RL technique to solve the on-street parking 
problem. The proposed approach is applied to estimate the best routes in terms of 
short travel distance and large number of available roadside PSs, and the design allows
drivers to choose their preferred objectives. A moving window mechanism is also incorporated into the underlying RL algorithm to better
cope with non-stationary environments. 
Experimental results have proven the efficiency of our system in improving the parking problems. 
For future work, we plan to include more objectives into our optimisation problem (e.g. travel time, 
walking distance), and extend the range of allowable values of $\alpha$.


%% file: main.bbl
\begin{thebibliography}{10}

\bibitem{shoup2006cruising}
D.~C. Shoup, ``Cruising for parking,'' {\em {Transport Policy}}, vol.~13,
  no.~6, pp.~479--486, 2006.

\bibitem{fog}
C.~{Tang} {\em et~al.}, ``{Towards Smart Parking Based on Fog Computing},''
  {\em IEEE Access}, vol.~6, pp.~70172--70185, 2018.

\bibitem{ict4cart2020}
M.~Buchholz {\em et~al.}, ``{Enabling automated driving by ICT infrastructure:
  A reference architecture},'' in {\em Proceedings of TRA'20, Helsinki
  (conference cancelled)}, 2020.
\newblock doi: 10.18725/OPARU-26023.

\bibitem{sptoken_journal}
R.~{Overko} {\em et~al.}, ``{Spatial Positioning Token (SPToken) for Smart
  Mobility},'' {\em IEEE Transactions on Intelligent Transportation Systems},
  pp.~1--14, 2020.

\bibitem{iccve}
R.~{Overko} {\em et~al.}, ``{Spatial Positioning Token (SPToken) for Smart
  Mobility},'' in {\em Proceedings of IEEE ICCVE'19}, pp.~1--6, 2019.

\bibitem{garlichs2020leveraging}
K.~Garlichs {\em et~al.}, ``{Leveraging the Collective Perception Service for
  CAM Information Aggregation at Intersections},'' 2020.
\newblock Accepted at IEEE VTC2020-Fall.

\bibitem{blockchain_parking}
W.~Al~Amiri {\em et~al.}, ``{Privacy-Preserving Smart Parking System Using
  Blockchain and Private Information Retrieval},'' in {\em Proceedings of
  SmartNets'19}, pp.~1--6, 2019.

\bibitem{p_span}
J.~{Ni} {\em et~al.}, ``{Privacy-Preserving Smart Parking Navigation Supporting
  Efficient Driving Guidance Retrieval},'' {\em IEEE Transactions on Vehicular
  Technology}, vol.~67, no.~7, pp.~6504--6517, 2018.

\bibitem{rl_lavp}
M.~{Khalid} {\em et~al.}, ``{A Reinforcement Learning based Path Guidance
  Scheme for Long-range Autonomous Valet Parking in Smart Cities},'' in {\em
  Proceedings of ComNet'20}, pp.~1--7, 2020.

\bibitem{non_stat_rl}
S.~Padakandla {\em et~al.}, ``{Reinforcement Learning in Non-Stationary
  Environments},'' {\em Applied Intelligence}, vol.~50, pp.~3590--3606, 2020.

\bibitem{wu2019early}
M.-C. Wu {\em et~al.}, ``Early detection of vacant parking spaces using dashcam
  videos,'' in {\em Proceedings of AAAI'19}, vol.~33, pp.~9613--9618, 2019.

\bibitem{bock2015street}
F.~Bock {\em et~al.}, ``On-street parking statistics using lidar mobile
  mapping,'' in {\em Proceedings of ITSC'15}, pp.~2812--2818, IEEE, 2015.

\bibitem{cdc}
R.~{Overko} {\em et~al.}, ``{Reinforcement Learning Augmented Optimization for
  Smart Mobility},'' in {\em Proceedings of IEEE CDC'19}, pp.~1286--1292, 2019.

\bibitem{pmlr-v80-zanette18a}
A.~Zanette {\em et~al.}, ``{P}roblem {D}ependent {R}einforcement {L}earning
  {B}ounds {W}hich {C}an {I}dentify {B}andit {S}tructure in {MDP}s,'' in {\em
  Proceedings of MLR'18}, vol.~80, pp.~5747--5755, PMLR, 10--15 Jul 2018.

\bibitem{ubev}
C.~Dann {\em et~al.}, ``Unifying {PAC} and regret: {U}niform {PAC} bounds for
  episodic reinforcement learning,'' in {\em Proceedings of NeurIPS'17},
  pp.~5713--5723, 2017.

\bibitem{sumo}
P.~A. Lopez {\em et~al.}, ``{Microscopic Traffic Simulation using SUMO},'' in
  {\em Proceedings of ITSC'18}, pp.~2575--2582, IEEE, 2018.

\end{thebibliography}
